\documentclass[sigconf,10pt]{acmart}

\usepackage{textcomp}
\usepackage{pifont}
\usepackage{xspace}

\usepackage{nicefrac}
\usepackage[utf8]{inputenc}
\usepackage[english]{babel}
\PassOptionsToPackage{bookmarks=false}{hyperref}
\usepackage{multirow, multicol, booktabs, tabulary, tabu, longtable, array, varwidth}
\usepackage{placeins, lipsum}
\usepackage[flushleft]{threeparttable}
\usepackage{tabularx}
\usepackage{amsmath}
\usepackage{pifont}
\usepackage{anyfontsize}

\usepackage{makecell}

\hypersetup{breaklinks=true}
\usepackage[labelfont=bf]{caption}
\usepackage{textcomp}
\usepackage[shortcuts,acronym]{glossaries}
\usepackage{soul, footnote, xargs}

\usepackage{amsmath}
\usepackage[utf8]{inputenc}
\usepackage{enumerate}

\usepackage[inline]{enumitem}
\setlist{noitemsep,nolistsep,leftmargin=*}

\usepackage{amsmath, amsfonts, amsthm, nicefrac}
\theoremstyle{definition}
\newtheorem{definition}{Definition}[section]

\usepackage{listings}

\usepackage[scaled=0.82]{beramono}

\usepackage[]{cleveref} % get fancy referencing
\crefname{appsec}{Appendix}{Appendices}
\crefformat{definition}{Def. #2#1#3}
\crefformat{section}{Section #2#1#3}
\crefformat{subsection}{Section #2#1#3}
\crefformat{subsubsection}{Section #2#1#3}
\crefformat{equation}{(#2#1#3)}
\crefrangeformat{equation}{(#3#1#4--#5#2#6)}
\crefmultiformat{equation}{(#2#1#3)}{ and~(#2#1#3)}{, (#2#1#3)}{ and~(#2#1#3)}
\crefformat{figure}{Fig.~#2#1#3}
\crefrangeformat{figure}{Figs. #3#1#4--#5#2#6}
\crefmultiformat{figure}{Figs.~#2#1#3}{ and~#2#1#3}{, #2#1#3}{ and~#2#1#3}
\crefname{algocf}{Alg.}{Algs.}
\Crefname{algocf}{Algorithm}{Algorithms}
\crefformat{table}{Table~#2#1#3}
\crefrangeformat{table}{Tables~#3#1#4--#5#2#6}
\crefmultiformat{table}{Tables~#2#1#3}{ and~#2#1#3}{, #2#1#3}{ and~#2#1#3}

\usepackage{textcomp}
\usepackage[shortcuts,acronym]{glossaries}
\usepackage{soul, footnote, xargs}
\usepackage{balance}

\usepackage{graphicx}
\usepackage{tikz, epstopdf, stfloats, bbding, capt-of}
\usepackage{algorithm}% http://ctan.org/pkg/algorithms
\usepackage{algpseudocode}% http://ctan.org/pkg/algorithmicx

\definecolor{Maroon}{RGB}{192,0,0}
\usepackage{wasysym}

\newcommand*\circled[1]{\tikz[baseline=(char.base)]{
        \node[shape=circle,draw,inner sep=1pt, Maroon, fill=Maroon] (char)
              {\color{white}\scriptsize\textbf{#1}};}%
        }

\newcommand{\BLUE}[1]{{\color{black} #1}}

\newcommand{\cmark}{\ding{51}}%
\newcommand{\xmark}{\ding{55}}%

\newcommand{\sysname}{{QLM}\xspace}

\copyrightyear{2024}
\acmYear{2024} 
\setcopyright{rightsretained}
\acmConference[SoCC '24]{ACM Symposium on Cloud Computing}{November 20--22,2024}{Redmond, WA, USA}
\acmBooktitle{ACM Symposium on Cloud Computing (SoCC '24), November 20--22, 2024, Redmond, WA, USA}
\acmDOI{10.1145/3698038.3698523}
\acmISBN{979-8-4007-1286-9/24/11}
\begin{document}

\title{Queue Management for SLO-Oriented Large Language Model Serving}

\author{Archit Patke}
\affiliation{%
  \institution{University of Illinois at Urbana-Champaign}
  \city{Urbana}
  \state{Illinois}
  \country{USA}
}
\email{apatke@illinois.edu}

\author{Dhemath Reddy}
\affiliation{%
  \institution{University of Illinois at Urbana-Champaign}
  \city{Urbana}
  \state{Illinois}
  \country{USA}
}
\email{dhemath2@illinois.edu}

\author{Saurabh Jha}
\affiliation{%
  \institution{IBM Research}
  \city{Yorktown Heights}
  \state{New York}
  \country{USA}
}
\email{Saurabh.Jha@ibm.com}

\author{Haoran Qiu}
\affiliation{%
  \institution{University of Illinois at Urbana-Champaign}
  \city{Urbana}
  \state{Illinois}
  \country{USA}
}
\email{haoranq4@illinois.edu}

\author{Christian Pinto}
\affiliation{%
  \institution{IBM Research}
  \city{Dublin}
  % \state{Ireland}
  \country{Ireland}
}
\email{Christian.Pinto@ibm.com}

\author{Chandra Narayanaswami}
\affiliation{%
  \institution{IBM Research}
  \city{Yorktown Heights}
  \state{New York}
  \country{USA}
}
\email{chandras@us.ibm.com}

\author{Zbigniew Kalbarczyk}
\affiliation{%
  \institution{University of Illinois at Urbana-Champaign}
  \city{Urbana}
  \state{Illinois}
  \country{USA}
}
\email{kalbarcz@illinois.edu}

\author{Ravishankar Iyer}
\affiliation{%
  \institution{University of Illinois at Urbana-Champaign}
  \city{Urbana}
  \state{Illinois}
  \country{USA}
}
\email{rkiyer@illinois.edu}



\renewcommand{\shortauthors}{A. Patke et al.}

\begin{abstract}
Large language model (LLM) serving is becoming an increasingly critical workload for cloud providers. 
Existing LLM serving systems focus on interactive requests, such as chatbots and coding assistants, with tight latency SLO requirements.
However, when such systems execute batch requests that have relaxed SLOs along with interactive requests, it leads to poor multiplexing and inefficient resource utilization.
To address these challenges, we propose \sysname, a queue management system for LLM serving.
\sysname maintains batch and interactive requests across different models and SLOs in a \emph{request queue}.
Optimal ordering of the request queue is critical to maintain SLOs while ensuring high resource utilization.
To generate this optimal ordering, \sysname uses a Request Waiting Time (RWT) Estimator that estimates the waiting times for requests in the request queue.
These estimates are used by a global scheduler to orchestrate LLM Serving Operations (LSOs) such as request pulling, request eviction, load balancing, and model swapping.
Evaluation on heterogeneous GPU devices and models with real-world LLM serving dataset shows that \sysname improves SLO attainment by 40--90\% and throughput by 20--400\% while maintaining or improving device utilization compared to other state-of-the-art LLM serving systems.
\sysname's evaluation is based on the production requirements of a cloud provider.
\sysname is publicly available at \href{https://www.github.com/QLM-project/QLM}{https://www.github.com/QLM-project/QLM}.
\end{abstract}

\maketitle

\section{Introduction}
\label{s:introduction}

% Why is queueing a problem? current queues cannot handle it, why use LSOs with queues? why merge LSOs with queuing

\noindent \textbf{Motivation.} 
Large language models (LLMs) such as OpenAI GPT-4 and Google Gemini have enabled novel capabilities in a wide range of AI applications~\cite{compound-ai-blog,bommasani2021opportunities,wu2023autogen} such as chatbots and coding assistants.
These base models such as GPT4 are further fine-tuned to support specialized tasks such as copywriting, financial planning, etc.~\cite{finetuning}.
Consequently, serving multiple models for enterprise and consumer applications with latency-oriented service-level objectives (SLOs) has become increasingly critical~\cite{jyothi2016morpheus,qiu2020firm,wang2024efficient}. 
% In addtion, production serving of LLMs involves handling requests targeting multiple models to be served on the same infrastructure due to the high cost of GPUs.

%Some focus on interactive services while others are batch jobs 
%Existing systems primarily focus on interactive jobs and do a great job with them
%However applying the same principles for batch jobs is bad because of high resource usage
%Instead, we propose the system: QLM where you can use request queues to queue up batch jobs which opens scope for reducing device requirement further by:
%1. allowing high priority requests to share the same GPUs as batch jobs
%2. allow multiple models to swap within the same LLM serving instance to minimize total GPU requirements
%One critical question becomes: while these mechanisms exist, how can they be orchestrated to maintain SLOs for batch and latency-critical jobs?

Previous work in this area~\cite{kwon2023efficient,yu2022orca,tensorrt,tgi,aiops2024qiu,li2023alpaserve,sheng2023flexgen,wu2023fast} largely focused on serving interactive requests, such as chatbots, with tight latency SLO requirements.
However, the recent explosive growth of LLM  applications, has generated a need to support batch LLM queries with SLO values ranging from minutes to hours for tasks such as data wrangling~\cite{narayan2022can}, document processing~\cite{jin2024comprehensive}, and model fine-tuning~\cite{sudalairaj2024lab}.

Given the broader range of  SLO requirements and the use of multiple models, maintaining \emph{request queues} is beneficial. 
Conceptually, when resources are limited, requests with relaxed SLOs can be kept at the back of the request queue and do not need to be executed immediately, while requests with tight SLOs can be kept ahead in the queue to ensure immediate execution and prevent \emph{head-of-line (HOL) blocking}.
Hence, queue ordering becomes an important decision-making problem for SLO-oriented LLM serving.

Previous work in SLO-oriented serving has largely focused on traditional DNN serving workloads such as CNNs and RNNs~\cite{resnet,salehinejad2017recent}, where the queue ordering decisions are made by systems such as Clockwork~\cite{gujarati2020serving}, INFaaS~\cite{romero2021infaas}, and SHEPHERD~\cite{zhang2023shepherd}.
Such systems leverage the deterministic execution times of DNN workloads to estimate queuing times and enable optimal ordering decisions that would maintain request SLOs.
However, such systems cannot be easily applied to LLM serving workloads because the execution time per request is non-deterministic as the number of output tokens are unknown apriori.
Using these system's assumption of fixed size batches with deterministic execution times for scheduling decisions leads to sub-optimal scenarios, described below.

% However, maintaining request queues also leads to \emph{head-of-line (HOL) blocking} and SLO violations.

\begin{figure}[!tb]
    \centering
        \hspace{-3ex}
    \begin{minipage}{.1\textwidth}
\includegraphics{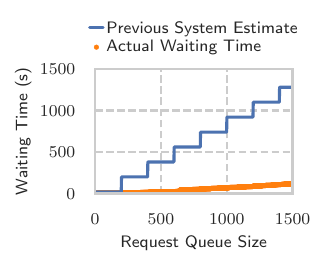}
    \end{minipage}
    \hspace{22ex}
    \begin{minipage}{.2\textwidth}
    \includegraphics{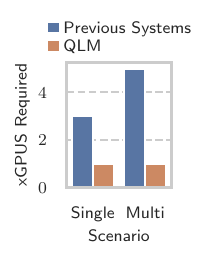}
    \end{minipage}
    \caption{Previously proposed SLO-oriented serving systems overestimate queue waiting time leading to suboptimal resource usage. (Left) Estimated waiting time when requests are run with Llama-70B on A100 GPUs with vLLM. (Right) Number of GPUs required to maintain 20s time-to-first-token (TTFT) SLO with previous systems vs \sysname in single and multi-model scenarios.}
    \Description{Previously proposed SLO-oriented serving systems overestimate queue waiting time leading to suboptimal resource usage. (Left) Estimated waiting time when requests are run with Llama-70B on A100 GPUs with vLLM. (Right) Number of GPUs required to maintain 20s time-to-first-token (TTFT) SLO with previous systems vs \sysname in single and multi-model scenarios.}
    \label{fig:waiting_time_misestimation}
\end{figure}

\begin{figure}[!tb]
    \centering
    \includegraphics[width=\linewidth]{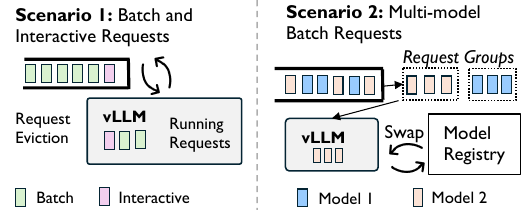}
    \caption{\sysname uses request groups and LLM Serving Operations (LSOs) such as request eviction to minimize resource requirement. Previously proposed systems would use four vLLM instances (compared to two for \sysname) due to limitations described in Figure~\ref{fig:waiting_time_misestimation}.}
    \Description{\sysname uses request groups and LLM Serving Operations (LSOs) such as request eviction to minimize resource requirement. Previously proposed systems would use four vLLM instances (compared to two for \sysname) due to limitations described in Figure~\ref{fig:waiting_time_misestimation}.}
    \label{fig:scenarios}
\end{figure}

First, previously proposed serving systems are unable to effectively multiplex interactive and batch requests on the same serving instance.
We find that the queue waiting time in LLM serving is lower than estimate used by the scheduler in such systems as shown in Figure~\ref{fig:waiting_time_misestimation} (left).  
As these systems generate a higher than actual estimate of the total queuing time, they allocate batch and interactive requests on separate serving instances, even in scenarios where allocation on the same instance would suffice to meet SLOs.

Second, previously proposed systems are unable to effectively multiplex different models on the same serving instance.
Similar to the first scenario, systems such as Clockwork~\cite{gujarati2020serving} overestimate the time required to serve a single model and prefer to allocate independent instances for each model to preserve SLOs.
Others such as SHEPHERD~\cite{zhang2023shepherd}, altogether forgo serving multiple models on the same instance due to the high cost of model swapping relative to inference times.

\noindent \textbf{Our Work.}
To address these limitations, we propose \sysname, a queue management system for LLM serving that maximizes SLO attainment while maintaining high throughput.
At the core of \sysname is the \emph{Request Waiting Time (RWT) Estimator} (described in Section~\ref{s:base_estimator}) that estimates the waiting times for requests in the request queue.
We observe that as the queue size grows larger, statistical effects of continuous batching~\cite{yu2022orca} in LLM serving allows us to create a tighter bound on waiting time.
Additionally, we empirically validate and provide a proof for applicability of the RWT estimator.
\sysname leverages this improved estimation to allow for better utilization and decrease resource cost by making closer to optimal request queue reordering decisions.

To order requests in the queue, \sysname, similar to SHEPHERD, uses the abstraction of \emph{request groups} (described in Section~\ref{s:virtual_queues}).
Each request group is a collection of requests that have relatively homogeneous performance requirements (e.g. similar SLOs and model).
\sysname places these request groups in \emph{virtual queues} that determine the order in which requests are consumed by the LLM serving instances.
In the LLM serving context, in addition to the benefits of scalability and predictability found in DNN serving, request groups are a useful abstraction for model swapping.
As decision making is made at the per-request group level, the total amount of swapping is minimized and throughput is improved as depicted in Figure~\ref{fig:scenarios} (Scenario 2).

Finally, \sysname uses a \emph{Global Scheduler} (described in Section~\ref{s:plan_generation}) that utilizes the waiting time estimates of request groups to create an optimal ordering and assignment of request groups onto the virtual queues to maximize SLO attainment.
The virtual queue ordering is used by four basic LLM Serving Operations (LSOs) to manage the request queue (see \cref{s:lso} for details):
\begin{enumerate*}[label=(\arabic*)]
\item \emph{Request Pulling} from the global waiting queue into the running batch in the GPU,
\item \emph{Request Eviction} from the running batch back into the waiting queue,
\item \emph{Model Swapping} from CPU to GPU memory, and
\item \emph{Load Balancing} across multiple LLM model instances.
For example, request eviction allows serving batch and interactive requests on the same LLM serving instance and prevents HOL blocking for interactive requests as shown in Figure~\ref{fig:scenarios} (Scenario 1).
\end{enumerate*}

\noindent \textbf{Results.}
We demonstrate \sysname on vLLM~\cite{kwon2023efficient} as the backend LLM-serving system.
We evaluate \sysname on three popular LLMs of varying sizes (i.e., Mistral-7B~\cite{mistral}, Vicuna-13B~\cite{vicuna}, and Llama-70B~\cite{touvron2023llama}) on GPU clusters with NVIDIA A10 and A100 GPUs.
We adopt workloads from a real-world LLM dataset: ShareGPT~\cite{shareGPT} using setups derived from our production requirements.
Our experiments demonstrate the following major improvements with \sysname:
\begin{enumerate}[wide=0pt,label=(\arabic*)]
    \item \emph{SLO Attainment:} Depending on the arrival rate, \sysname achieves 40--90\% higher SLO attainment compared to the vanilla vLLM serving system and 50--90\% higher SLO attainment compared to traditional ML serving systems like SHEPHERD.
    \item \emph{Request Throughput:} \sysname improves the request throughput in a multi-model serving system by 400\% on average and in a single-model serving system by 20\% on average compared to other LLM serving systems.
    \item \emph{LSO Ablation Study:} \sysname demonstrates that all LSOs contribute to SLO attainment and throughput improvement.
    Notably, we find that model swapping improves throughput by 300\% in multi-model serving, and request eviction improves SLO attainment by 80\% in single-model serving.
\end{enumerate}
\sysname has been merged into an internal production LLM routing service. 
\section{Background}
\label{s:background}

\subsection{LLM Inference}
\noindent
\textbf{Inference Primer.}
An inference process starts from a request (prompt) with a list of input tokens \((x_1, \ldots, x_n)\).
The LLM generates a list of output tokens \((x_{n+1}, \ldots, x_{n+T})\).
% according to Eq. (1).
Due to the \textit{autoregressive} pattern, the LLM can only generate new tokens one by one, and the generation process of each new token depends on all the previous tokens in that sequence, specifically their key and value vectors.
In this sequential generation process, the key and value vectors of existing tokens are cached for generating future tokens, known as \textit{KV cache}.

Therefore, given a LLM request prompt, the generation computation can be decomposed into two phases:
\begin{enumerate*}[label=(\arabic*)]
    \item A \textit{prefill} stage takes the whole user prompt \((x_1, \ldots, x_n)\) as input and computes the probability of the first output token \(P(x_{n+1} | x_1, \ldots, x_n)\).
    % During this process, also generates the key vectors \(k_1, \ldots, k_n\) and value vectors \(v_1, \ldots, v_n\). 
    \item A \emph{decoding stage} (autoregressive generation) generates the remaining output tokens sequentially.
    At iteration \(t\), the model takes one token \(x_{n+t}\) as input and computes the probability \(P(x_{n+t+1} | x_1, \ldots, x_{n+t})\) with the key vectors \(k_1, \ldots, k_{n+t}\) and value vectors \(v_1, \ldots, v_{n+t}\). This phase completes when an end-of-sequence (\texttt{<eos>}) token is emitted.
\end{enumerate*}

\noindent
\textbf{Continuous Batching.}
During LLM inference, the decoding stage is memory-bound, as loading model weights from memory takes longer than computation.
Therefore, state-of-the-art LLM serving systems like vLLM~\cite{kwon2023efficient}, Orca~\cite{yu2022orca}, Tensor-RT~\cite{tensorrt} and TGI~\cite{tgi} employ continuous batching with iterative scheduling to enable dynamic addition of requests to a batch as soon as others have finished generation.
% Compared to fixed batch sizes, these systems improve memory efficiency by processing multiple requests simultaneously.

\noindent
\textbf{PagedAttention.}
Static allocation of the KV cache can result in significant memory waste as the KV cache grows dynamically during the decoding stage.
PagedAttention~\cite{kwon2023efficient} introduces the idea of managing the KV cache, like OS memory, via pages and enabling dynamic allocation.
Such dynamic allocation prevents fragmentation and enables nearly 100\% utilization of GPU memory and furthers throughput improvement when combined with continuous batching~\cite{kwon2023efficient}.

% \begin{figure}[!tb]
%     \centering
%     \includegraphics[width=\linewidth]{resources/motivation_diagram.pdf}
%     \caption{Interaction between LLM applications and a multi-model LLM serving system.}
%     \label{fig:llm_serving_system}
% \end{figure}
\begin{figure*}[!ht]
    \centering
    \begin{minipage}{.28\textwidth}
        \centering    \includegraphics{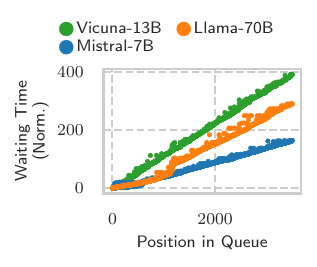}
        \vspace*{-20pt}
        \caption{Requests have predictable waiting times in a continuous batching system.}
        \Description{Requests have predictable waiting times in a continuous batching system.}
        \label{fig:Predictable Waiting Time}
    \end{minipage}%
    \hfill
    \centering    
        \begin{minipage}{.4\textwidth}
        \centering
        \includegraphics{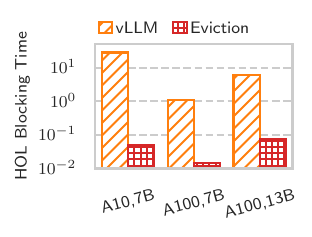}
        \caption{Forced request eviction leads to reduction in head-of-line (HOL) blocking time.}
        \Description{Forced request eviction leads to reduction in head-of-line (HOL) blocking time.}
        \label{fig:hol_blocking_time}
    \end{minipage}%
    \hfill
    \centering    
    \begin{minipage}{.28\textwidth}
    \centering
    \includegraphics{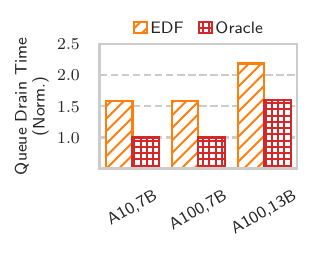}
    \caption{Model swapping and request pulling can jointly decrease queue drain time.}
    \Description{Model swapping and request pulling can jointly decrease queue drain time.}
    \label{fig:swaping_with_reordering}
    \end{minipage}%
    \hfill
\end{figure*}

\subsection{LLM Serving Systems}

% \BLUE{Figure~\ref{fig:llm_serving_system} shows the typical workflow of an LLM serving system.
\BLUE{
User-facing applications such as chatbots, log processing, and recommenders have specific latency SLO requirements~\cite{qiu2020firm,jyothi2016morpheus,qiu2024atc,chen2019parties}.
When interacting with LLMs, each application generates requests that consist of the \textit{input prompts} and associated \textit{metadata} (e.g., model type and SLO value) to the LLM serving system.
For example, chatbots require requests to complete by a deadline (e.g., p99 time to first token (TTFT) < 20s~\cite{gnewuch2022opposing}), and batch jobs like document processing have a more relaxed SLO in the order of minutes to hours.
Requests may also need to be served by multiple fine-tuned models specialized for various tasks.
For example, Code Llama~\cite{roziere2023code} is fine-tuned for coding assistance, and Llama-chat~\cite{touvron2023llama} is fine-tuned for chatbots.
% These requests are then served by an LLM serving system such as vLLM~\cite{kwon2023efficient}, TGI~\cite{tgi}, and Triton~\cite{triton} that implement state-of-the-art inference optimizations such as continuous batching and PagedAttention.
Maintaining a standalone LLM serving system for each of these models or SLO types can be expensive, and they often have to be multiplexed together to share the same serving system~\cite{choi2022serving,li2023alpaserve,lemay2020perseus,sheng2023slora}.
Multiplexing batch and interactive requests across various models with limited resources leads to formation of request queues, and managing these queues is critical to comply with SLO requirements.}

% However, even though these systems are optimized for high throughput serving, served requests can suffer from \emph{head-of-line (HOL) blocking}~\cite{viernickel2018multipath,aiops2024qiu} when the request arrival rate exceeds the LLM serving system throughput~\cite{wang2024efficient}.

% LLM serving systems have to manage these requests and models with limited devices through various backend LLM serving operations (LSOs) as defined in \cref{def:lso} such as request eviction~\cite{wu2024loongserve}, load balancing~\cite{griggs2024m}, and GPU-CPU state swap~\cite{strati2024dejavu}, in order to meet the SLO requirements.
% \emph{Managing these LSOs can be complex because of
% \begin{enumerate*}[label=(\arabic*)]
%     \item their inter-dependency with each other, and
%     \item the auto-regressive nature of LLMs where the length of output tokens is highly variable.
% \end{enumerate*}}

\subsection{\sysname Definitions}
Before we present the motivation and characterization study, below are definitions of the terms that we use.

\begin{definition}
    \emph{Request:} Each request consists of the prompt (i.e., input tokens) and its associated metadata (e.g., model type).
    Requests arrive at varying rates and burstiness, leading to request queues with dynamically changing sizes.
\end{definition}

\begin{definition}
    \emph{SLO:} Each request arrives with a service-level objective (SLO) value that enforces the \emph{time to first token (TTFT)} for the request.
    While \sysname primarily focuses on TTFT, it can be paired with another system such as Andes~\cite{liu2024andes} to also maintain inter-token latency (ITL).
\end{definition}

\begin{definition}
\label{def:serving-instance}
    \emph{LLM Serving Instance:} An LLM serving system\footnote{The difference between an LLM serving system and an LLM serving instance is similar to JAVA classes and objects. For example, vLLM is an LLM serving system while vLLM with a loaded model like Llama 70B is an instance of the LLM serving system.} is capable of hosting LLM models by providing the necessary infrastructure and resources to load the models into memory and respond to requests.
    \sysname is compatible with existing LLM serving systems such as vLLM~\cite{kwon2023efficient} and TGI~\cite{tgi}.
    An LLM serving instance is composed of the LLM serving system and an LLM model that is being served.
\end{definition}

\subsection{Motivation and Characterization}
\label{ss:characterization}

To enable SLO-oriented LLM serving, it is critical to understand (1) the impact of LLM autoregressive patterns on the request waiting time, (2) the performance implications of queuing batch and interactive requests, and (3) the performance implications of queuing multi-model batch requests.

We characterize these scenarios with a state-of-the-art LLM serving system, vLLM~\cite{kwon2023efficient}, to motivate the design of \sysname.
We use ShareGPT~\cite{shareGPT} traces to evaluate the system.
We present three key insights below.

\noindent \textbf{Insight \#1: \textit{Waiting times in long request queues can be accurately estimated analytically.}}
Recall from Figure~\ref{fig:waiting_time_misestimation} that incorrectly estimating waiting times can lead to  resource waste.
Therefore, we attempt to build an accurate queue waiting time estimator.

While the completion time for individual requests in LLM inference can vary widely, the average waiting time for a request in a long queue is predictable. 
Due to the statistical averaging effects resulting from a large number of requests (as detailed in Appendix~\ref{s:waiting_time_proof}), the waiting time can be estimated by dividing the total number of output tokens for the pending requests by the token generation throughput.
The total number of output tokens adheres to a Normal distribution, in accordance with the Central Limit Theorem, since individual requests are independent of each other

Figure~\ref{fig:Predictable Waiting Time} illustrates this linear relationship between waiting time and queue position when serving requests for three varying-sized LLMs on NVIDIA A100 GPUs.
Additionally, we find that the estimator is highly accurate with a coefficient of determination ($R^2$) of 0.99 (out of 1.0).

%Multiple LSOs needs to be connected back to SLOs and queues
\noindent \textbf{Insight \#2: \textit{HOL blocking times due to continuous batching can be in the order of tens of seconds.}}
The straightforward way to prioritize interactive requests over batch requests on the same LLM serving instance is by placing them at the front of the waiting queue.
However, request placement in the waiting queue may not be sufficient for immediate execution due to the lack of available GPU memory, causing head-of-line (HOL) blocking.
Therefore, in such scenarios, evicting batch requests from the GPU is required.
To minimize the cost of eviction, we can preserve the KV cache of batch requests, allowing execution to resume from the last decoding iteration.

Figure~\ref{fig:hol_blocking_time} illustrates the HOL blocking time when run with a mixed workload comprising interactive and batch requests.
In the absence of request eviction, the HOL blocking time can be in the order of several seconds, which can lead to violation of latency SLOs for interactive requests. 
However, request eviction significantly reduces the waiting time because interactive requests only need to wait for a single decoding iteration before they can be scheduled, resulting in a 100--1000$\times$ reduction in waiting time.

\noindent \textbf{Insight \#3: \textit{Policies such as Earliest Deadline First (EDF) are insufficient to eliminate HOL blocking from model swapping.}}
The optimal request pulling strategy to maximize the number of requests that satisfy SLOs is Earliest Deadline First (EDF) scheduling.
\textit{However, this assumes that the model swapping cost is negligible.}
Frequent model swaps can happen (similar to thrashing) if multiple models are served to time share the same GPU devices, leading to SLO violations due to longer completion times to drain the queue and a drop in throughput.
For example, consider the case illustrated in Figure~\ref{fig:swaping_with_reordering}.
Requests with varying SLOs arrive in the queue, and they are placed by an EDF policy, causing multiple model swaps and substantially higher time to drain the entire request queue.
Specifically, we find that across models and GPUs, the time required to serve all requests in the queue (i.e., the queue drain time) is substantially higher for the EDF policy compared to an Oracle policy that groups requests from the same model together to prevent the overhead of repetitive model swaps. 
\section{\sysname Design Overview}
\label{s:design}

\begin{figure}[ht]
    \centering
    \includegraphics{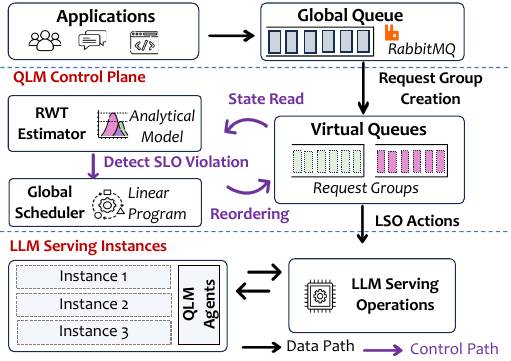}
    \caption{Overview of \sysname.}
    \Description{Overview of \sysname.}
    \label{fig:qlm_overview}
\end{figure}

\sysname aims to maximize latency SLO satisfaction in LLM serving workloads.
To do so, \sysname manages a global request queue and orchestrates multiple LLM Serving Operations (LSOs) to reorder and drain the global queue.

\subsection{Lifecycle of a Request in \sysname}
To explain the design of \sysname (as shown in Figure~\ref{fig:qlm_overview}), we first walk through the lifecycle of a request generated from applications to completion.
LLM-augmented applications generate requests that are received at the \sysname API gateway.
These requests are added to a global queue where they wait until being served.
To prevent the global queue from being a single point of failure, it is implemented with a distributed message broker such as RabbitMQ~\cite{rabbitmq} that provides the requisite fault tolerance and consistency properties.

% lifecycle step 1
\noindent
\textbf{Formation of Request Groups.}
Every incoming request is grouped with other requests that share common performance characteristics (such as model type, SLO value, and token distribution) to form \textit{Request Groups}.
This converts the complexity of the optimization problem from per-request level to per-request-group level.
By doing so, it alleviates the scalability challenges and lowers optimization overheads.
Additionally, request groups are a useful abstraction in the multi-model serving case as described in Section~\ref{ss:characterization}. 
Request grouping criteria and details are described in \cref{s:virtual_queues}.

% \noindent \textbf{Request Groups to Reduce Overhead.}
% As actions for multiple backend LSOs are set when serving requests from the global queue, \sysname needs to solve a multi-dimensional optimization problem that involves SLOs, model type, waiting time, etc.
% Solving such an optimization problem per-request is not scalable due to the high solving overhead.
% Hence, to reduce such overhead, \sysname uses the notion of \emph{request groups} that scales down the input space for the optimization problem.
% \sysname creates these request groups based on the performance demand/requirement characteristics of requests.
% We identify input/output tokens, SLOs, and model types as the necessary features to capture the entire performance spectrum.

% lifecycle step 2
\noindent
\textbf{Assigning Request Groups to Virtual Queues.}
Requests in a request group are then assigned to a \textit{Virtual Queue}, representing a waiting queue for an LLM serving instance in the cluster.
The introduction of virtual queues creates a common abstraction for setting the actions of backend LLM Serving Operations (LSOs) such as request pulling, request eviction, load balancing, and model swapping.
The ordering of the request groups in a virtual queue determines the execution ordering of the requests on the corresponding LLM serving instance.
We refer readers to \cref{s:virtual_queues} for virtual queue formation and \cref{s:lso} for the translation from virtual queue ordering to LSO actions.

% \noindent \textbf{Virtual Queues as an LSO Abstraction.}
% Additionally, we introduce the notion of \emph{virtual queues} that creates a common abstraction for setting the actions of underlying LSOs.
% A \emph{virtual queue} is an ordering of request groups that are served by an LLM serving instance.
% Multiple virtual queues are used by \sysname, and each virtual queue corresponds to its own LLM serving instance.
% The virtual queue ordering sets LSO actions using \sysname agents resident on each LLM serving instance.
% For example, to configure the request eviction LSO, changes in the head of the virtual queue are monitored by a \sysname agent.

% lifecycle step 3
\noindent
\textbf{Virtual Queue Reordering for SLO Attainment Maximization.}
While requests are assigned to request groups in a first-come-first-serve manner, request groups in a virtual queue are reordered to maximize the SLO attainment for all requests being served.
At the core of SLO attainment maximization are \sysname's \emph{request waiting time (RWT) estimator} (see \cref{s:base_estimator}) and \emph{global scheduler} (see \cref{s:plan_generation}).

% \noindent \textbf{RWT Estimator and global scheduler for SLO Translation.}
% \sysname's translation of SLOs to virtual queue ordering and consequently LSO actions is enabled by the \emph{request waiting time (RWT) estimator} and \emph{global scheduler}.
% The RWT estimator calculates completion times from the virtual queue and the global scheduler uses these estimates to reorder the virtual queues and satisfy SLOs.
% The \textit{RWT estimator} (described in Section~\ref{s:base_estimator}) estimates the time required to serve this newly added request based on its position in the virtual queue.
% If the estimation is higher than the actual SLO value, then an SLO violation is detected and the \textit{global scheduler} is invoked to re-order all the virtual queues in an attempt to satisfy the SLO value.
% The global scheduler does this reordering based on a stochastic programming solver (described in Section~\ref{s:plan_generation}).

\noindent
\textbf{Request Execution.}
Each request, when being moved to the head of the virtual queue, will be executed on the LLM serving instance, and the output will be returned to the application.
This completes the lifecycle of a request.

We illustrate the rationale of \sysname's design choices and workflow in the next section (\cref{ss:design-principles}).

\subsection{\sysname Design Principles}
\label{ss:design-principles}

We highlight the major design principles underpinning \sysname, derived from large-scale production LLM serving workload requirements at a major cloud provider.

\noindent \textbf{Design Principle \#1: \textit{Scaling to a high request arrival rate and burstiness.}}
\sysname must be able to handle a high volume of requests for SLO attainment without the overhead that compromises the serving throughput.
Existing model-serving frameworks that leverage optimization techniques such as linear programming have exponential or cubical complexity, which limits the scalability to larger workloads and longer queues.
\sysname instead introduces \textit{request groups} to reduce the input space of the optimization solvers, thus lowering the computational overhead and enabling scalability.

\noindent
\textbf{Design Principle \#2: \textit{Handling multiple LSOs with inter dependencies.}}
To attain model-serving latency SLOs, it is critical to translate the latency SLO to the appropriate backend LSO actions.
\sysname models these complex inter-relationships with a two-step approach.
First, \textit{virtual queues} enable the necessary abstraction to enable actions for multiple backend LSOs.
Second, the \textit{global scheduler} models the impact of ordering on multiple LSOs with a linear programming solver. 
\BLUE{We specifically prefer a linear programming solver over other optimization methods because it systematically considers various constraints introduced by multiple LSOs, SLO constraints, and waiting time estimates from the RWT estimator.
}

\noindent
\textbf{Design Principle \#3: \textit{Handling heterogeneous models and hardware device configurations.}}
LLM serving workloads consist of diverse model types with vastly different computational requirements, SLOs, and token length distributions.
Hardware device configurations are also heterogeneous in terms of computing power, GPU memory capacity, and GPU-CPU memory bandwidth.
To efficiently map LLM requests to the appropriate hardware resources, \sysname's global scheduler has to consider each device's computing power, memory capacity, and memory bandwidth.
The \textit{RWT estimator} estimates this impact of heterogeneity for the global scheduler.
The profiling costs for the RWT estimator are minimal, only a single batch run for a given combination of request group and GPU device is needed.
Hence, \sysname does not require significant training when adding new LLM models or GPU devices into the serving cluster.
\section{Request Groups and Virtual Queues}
\label{s:virtual_queues}

In this section, we describe the concept of \textit{virtual queues} and the process of classifying LLM requests into \textit{request groups} and assigning request groups to virtual queues.
% \sysname's virtual queues draw inspiration from virtual output queuing (VOQ)~\cite{mckeown1997tiny,tamir1988high}, a popular architecture used in network switches to address head-of-line blocking.

\begin{definition}
\label{def:request-group}
    \emph{Request Group:} Each request group is a collection of multiple requests that are relatively homogeneous, i.e., sharing similar performance demand or requirement characteristics.
    We identify that input/output token distributions, model type, and SLO values are sufficient for the RWT estimator (as explained in Section~\ref{s:base_estimator}).
\end{definition}

\begin{definition}
\emph{Virtual Queues:}
Each virtual queue is a sequence of request groups that denotes the relative order in which requests will be served.
There is a one-to-one mapping between an LLM serving instance and a virtual queue.
\end{definition}

By creating the abstraction of request groups and virtual queues, the ordering of request groups in a virtual queue allows \sysname to configure actions for multiple downstream LSOs to attain latency SLOs for the LLM-serving requests in a scalable manner (described in Section~\ref{s:lso}).

\begin{algorithm}[t]
\begin{algorithmic}[1]
\State $groups \gets kMeansClustering(requests)$
\For{$i \gets 1$ to $length(groups)$}
    \If{$groups[i].size() > avg\_batch\_size \times \delta$}
        \State $newGroups \gets groups[i].splitHalf()$
        \State $groups.append(newGroups)$
    \EndIf
\EndFor

\end{algorithmic}
\caption{Request Group Creation}
\label{algo:clustering}
\end{algorithm}

\noindent \textbf{Request Group Creation.}
Request groups are created in two steps: (i) clustering similar requests based on \cref{def:request-group}, and (ii) splitting large request groups.
Algorithm~\ref{algo:clustering} describes the request group creation process.
The parameters identified for the request grouping include model types, input/output token distribution, and SLO values.
Grouped requests based on such parameters exhibit predictable request completion time distribution (compared to that of each individual request) as explained in Section~\ref{s:base_estimator}.
Additionally, we also limit the size of each request group to a small multiple ($\delta$) of batch size.
We refer the reader to \cref{ss:robustness} for the trade-off analysis between overhead and decision-making granularity: (1) Larger request groups would decrease the number of request groups and thus the overhead of the global scheduler; (2) However, restricting the size of request groups is beneficial as it allows for more fine-grained decisions.
Since requests within a request group are relatively homogeneous, \sysname treats the ordering of the requests within a group using a first-come-first-serve (FCFS) policy.
Request groups are dequeued from the virtual queue when all requests complete execution.

\noindent \textbf{Handling New Incoming Requests.}
As new requests join the global queue, they are classified into the existing request groups, and the RWT estimator calculation is triggered to find out whether any SLOs are being violated.
Upon any SLO violation, the global scheduler is called to reorder the request groups in the virtual queues to maximize SLO attainment given the current states (estimations).
% The global scheduler is rerun only when the RWT estimator anticipates the SLOs are going to be violated in response to new requests being added to the queue, eliminating the overhead of repeated use.

\noindent
\textbf{Fault Tolerance in Queue Management.}
\sysname only stores a single replica of the requests and their metadata in the global queue, which avoids the need to maintain consistency between multiple queues.
The global queue is implemented using a distributed message queue broker such as RabbitMQ~\cite{rabbitmq} that provides the necessary replication, fault tolerance, and persistence mechanisms.
The data structure that implements virtual queues records orderings of subsets of requests in the global queue.
These virtual queues are implemented as lightweight data structures that maintain pointers or references to the actual requests stored in the global queue.
By using virtual queues, \sysname can achieve the following:
\begin{enumerate*}[label=(\arabic*)]
\item \emph{Fault Isolation}: If an LLM serving instance fails, only the corresponding virtual queue is affected, and the remaining virtual queues can continue processing requests without interruption.
Request groups from the lost virtual queue are assigned to other virtual queues using the global scheduler.
\item \emph{Consistency}: Since the actual requests are stored in the global queue, virtual queues can be reconstructed or reassigned without compromising the consistency of the request data.
\end{enumerate*}
\section{LLM Serving Operations}
\label{s:lso}

The LLM serving instances serve requests from the corresponding virtual queue and execute backend LSO actions when necessary.
The LSOs by themselves are merely action actuators, and the intelligence required to configure when and which action to set comes from the virtual queue ordering set by the global scheduler (as described in \cref{s:plan_generation}).

\begin{figure}[!tb]
    \centering
    \includegraphics[width=\linewidth]{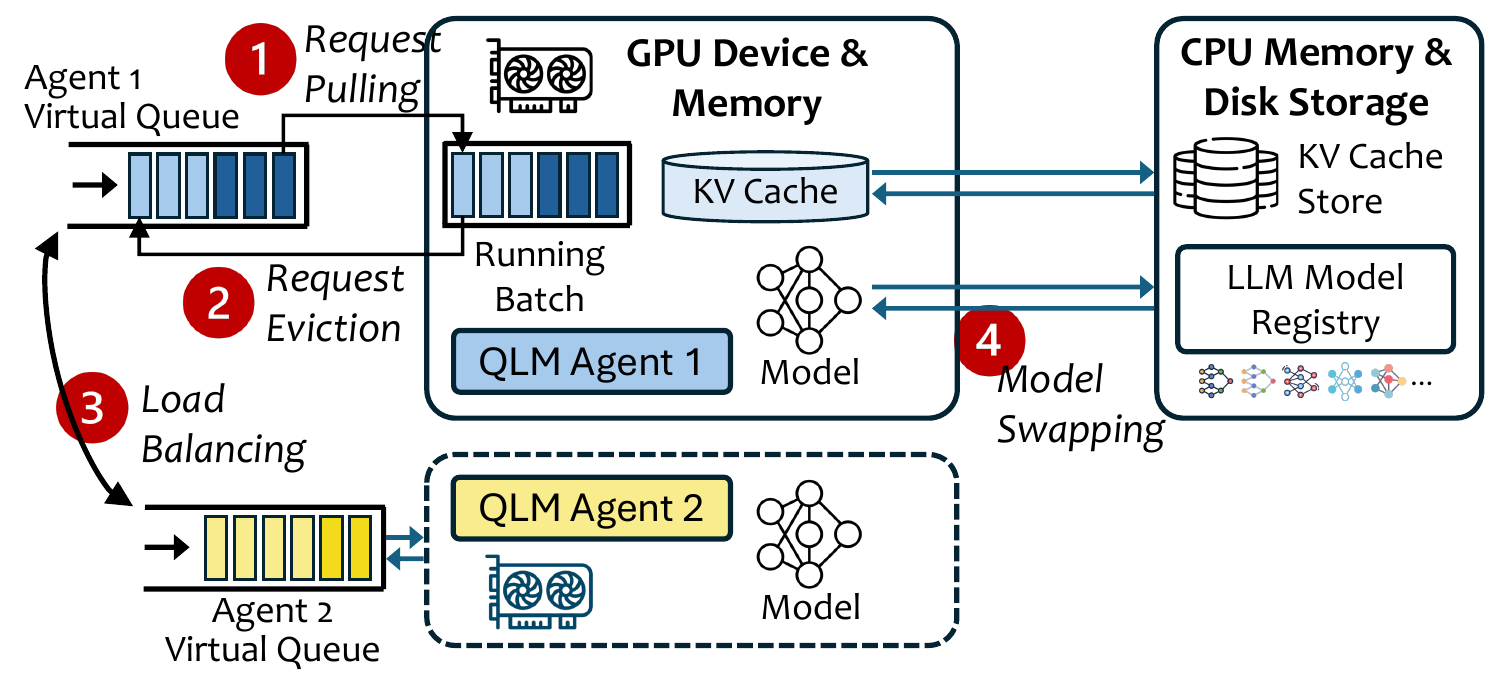}
    \caption{Basic LLM serving operations (LSOs) for an LLM-serving instance that a \sysname agent manages.}
    \Description{Basic LLM serving operations (LSOs) for an LLM-serving instance that a \sysname agent manages.}
    \label{fig:roms}
\end{figure}

\cref{fig:roms} shows the four basic LSOs that \sysname currently supports.
A \sysname agent resident on each LLM serving instance monitors the virtual queue ordering and converts it into LSO actions.
When the virtual queue state changes or new requests are added, \sysname agents initiate request pulling and load balancing.
Similarly, when the head request group changes, the \sysname agent initiates request eviction.
Model swapping is initiated by the \sysname agent for models at the head of the virtual queue.
Each of these LSOs modifies the internal state of the LLM serving instance, which includes the running batch of requests, KV cache store, and model weights.
Below, we describe each of these LSOs in detail and their action setup based on the virtual queue ordering.

\noindent \textbf{Request Pulling} (\circled{1} in \cref{fig:roms}).
Request pulling refers to the operation that dequeues the requests in the virtual queue using a pull-based model and adds it to the running batch, i.e., whenever the total tokens of the running batch are below the GPU capacity, a pull signal is issued to retrieve a request from the global queue.
The exact pulled request is determined by the request group at the head of the virtual queue.
Within the head request group, requests are ordered in an FCFS manner, therefore the first request to join the head request group would be the first to be dequeued from the virtual queue and added to the GPU's running batch.
Note that request pulling is insufficient to immediately serve a request with a low SLO value (i.e., stricter SLO) because a pull operation to the virtual queue can only happen if spare token capacity exists on the GPU device (i.e., without head-of-line blocking).

\noindent \textbf{Request Eviction} (\circled{2} in \cref{fig:roms}).
As request pulling by itself may not be sufficient to enable the immediate serving of requests with low SLO values due to head-of-the-line blocking, \sysname also supports request eviction.
Request eviction is invoked when the RWT estimator detects an SLO violation, and the global scheduler replaces an existing request group by placing a request group at the head of the virtual queue.
In request eviction, requests of the head request group are pulled into the running batch based on available capacity, and previously running requests are evicted (back) into the global queue.
\BLUE{To prevent re-computation of KV cache of the lost request upon eviction, we migrate it to CPU memory instead.}
However, the GPU-to-CPU memory bandwidth is typically at least 10$\times$ less than the GPU memory bandwidth, and if the evicted request has a large KV cache, it leads to significant transfer overhead and consequent performance degradation.
\sysname hides this performance degradation using the asynchronous GPU memory copy available in most GPU programming libraries.

\noindent \textbf{Load Balancing} (\circled{3} in \cref{fig:roms}).
As each LLM serving instance is associated with a separate virtual queue, the global scheduler's assignment of request groups to a virtual queue inherently performs load balancing.
Each instance would only pull from its associated virtual queue, thus ensuring the distribution of requests across all the serving instances.
Note that \sysname does not implement preemptive load balancing, i.e., once request groups start executing on an LLM serving instance, they cannot be migrated to another instance.

\noindent \textbf{Model Swapping} (\circled{4} in \cref{fig:roms}).
Each LLM serving instance can serve multiple models by switching the underlying model weights and flushing out the KV cache.
\sysname assumes a two-tier hierarchy of memory and disk storage.
Therefore, any model that needs to be served from the LLM model registry (located in the storage) has to undergo two distinct swaps:
\begin{enumerate*}[label=(\arabic*)]
    \item \emph{Storage-CPU swapping:} The model is first swapped from the LLM model registry to CPU memory, and
    \item \emph{CPU-GPU swapping:} The model located in the CPU memory is swapped into GPU memory for inferencing.
\end{enumerate*}
\sysname is able to decide the location (i.e. GPU mem, CPU mem or storage) of each model by checking the virtual queue order.
The model for the head request of the virtual queue is currently \textit{active} and should be placed in the GPU memory.
Models present later in the virtual queue are \textit{warm} and placed in the CPU memory until all the CPU memory is exhausted.
The remaining models (\textit{cold} models) are not swapped out from the LLM model registry (located in the storage).

\noindent \textbf{LSO Implementation.}
We implement the abovementioned LSOs on top of vLLM, a state-of-the-art LLM serving system.
\sysname agents are responsible for triggering each LSO action.
Request pulling and load balancing are implemented by async pull calls to the virtual queues when there is spare token capacity on the LLM serving instance.
Request eviction and GPU-CPU state swapping require instrumentation to the vLLM scheduler.
In each iteration, the vLLM scheduler attempts to generate a new token for all the running requests and preempts any request that exceeds the total GPU capacity.
At the end of an iteration, \sysname agent checks if either request eviction or state swapping is required and performs the operation.
For both operations, current requests are removed from the running batch to make space for the incoming requests at the head of the virtual queue.
For request eviction, the scheduler uses an asynchronous GPU transfer of the KV cache.
Model swapping into the GPU is implemented by changing the underlying model of the vLLM instance and flushing out the KV cache.
% \sysname also uses \texttt{vmtouch}~\cite{vmtouch} to ensure that models closer to the front of the queue (but not being actively used) are kept hot in CPU memory and not swapped out to the storage. 
\section{Request Waiting Time (RWT) Estimator}
\label{s:base_estimator}

\begin{table}[!t]
\centering
\caption{Glossary of symbols used in the RWT estimator.}
\resizebox{0.925\linewidth}{!}{%
\begin{tabular}{cl}
\toprule
\textbf{Symbol} & \textbf{Description} \\
\midrule
$C_{q}$ & Completion time for a request $q$ \\
$W_q$ & Request waiting time for a request $q$ \\
$P$ & Prefill time for a request \\
$D_{q}$ & Total decode time for a request $q$ \\
$O_q$ & Number of output tokens for a request $q$  \\
$\Theta$ & Token generation throughput \\
$\epsilon$ & Inefficiency factor due to continuous batching \\
$d$ & Decode time per output token \\
\bottomrule
\end{tabular}
}
\label{tab:symbols_rwt}
\end{table}

\noindent
\sysname leverages waiting time estimates from the RWT estimator for better utilization and decrease resource costs by making closer to optimal request queue reordering decisions.
The RWT estimator uses a statistical approach to generate estimates of waiting and completion times for requests.
\BLUE{To do so, the estimator first generates an upper bound on request completion time i.e. the actual request completion time would be lesser than the estimate.
These completion times are then aggregated to generate completion time of the entire request group.
Overall, the estimator is conservative when queue size is small (i.e. the estimate is higher than the actual waiting times) and the accuracy increases with increase in queue size.
}
% To create such a RWT estimator, we extend the estimator described in Section~\ref{ss:characterization} further by accounting for the variance per request in completion times.

The estimation process is explained in further detail below with variable definitions listed in~\cref{tab:symbols_rwt}.

As shown in Equation~\ref{eq:total_completion_time}, the total request completion time equals the sum of the waiting time ($W_q$), prefill time ($P$), and total decode time across all the output tokens ($D_q$) for a request $q$.
\begin{align}
C_{q} = W_q + P + D_{q}
\label{eq:total_completion_time}
\end{align}

\noindent
\textbf{Estimating Prefill Time.} The prefill time $P$ is typically constant per model type when the input tokens are small as it is a highly parallel GPU-accelerated operation whose time increases minimally as the number of input tokens increases~\footnote{Abnormally long context prompts beyond 10 standard deviations of production prompt distribution are not in the scope of \sysname and we leave it for future research work.}.
Experiments show that the latency increase from additional input tokens is 100$\times$ less compared to the latency increase from each additional output token~\cite{anyscale-blog}.
Therefore, the major distribution terms that still remain to be estimated are the waiting time ($W_q$) and decode time ($D_q$).

\noindent
\textbf{Estimating Waiting Time.}
We consider the token generation throughput ($\Theta$) to be constant throughout the token generation process due to statistical averaging effects described in Appendix~\ref{s:waiting_time_proof}.
Therefore, the total waiting time for a single request can be represented by Equation~\ref{eq:base_equation} by dividing the number of tokens ahead ($\sum_{i=1}^{q-1} O_i$) in the queue by the token generation throughput ($\Theta$) where $i$ denotes each of the $q-1$ requests in the queue ahead of the request we model.

\begin{align}
W_q = \sum_{i=1}^{q-1} \frac{O_i}{\Theta}
\label{eq:base_equation}
\end{align}

\noindent
Note that we do not know the number of output tokens ahead of time (that requires the knowledge of the output sequence for all requests in the waiting queue), so we model them as a distribution with the mean $\mu_o$ and standard deviation $\sigma_o$ fitted from the request input-output history dataset for the request group that the request $q$ belongs to.
\BLUE{As $q$ becomes larger, the Central Limit Theorem (CLT) applies and the assumption of Normal distribution is accurate for any underlying request output token distribution.
We further explain this in Appendix~\ref{s:waiting_time_proof}.}

\begin{align}
\sum_{i=1}^{q-1} O_i \sim N((q-1)\mu_o, (q-1)\sigma_o^2)
\label{eq:output_token_dist}
\end{align}

% Once we have the waiting time, we can derive the total completion time of the request by accounting for the token generation steps.

\noindent
\textbf{Estimating Decode Time.} We compute the total decode time using Equation~\ref{eq:total_decode_time}.
\begin{align}
D_q = O_q \times \epsilon \times d
\label{eq:total_decode_time}
\end{align}

\noindent

As we do not know the exact number of output tokens ($O_q$) in Equation~\ref{eq:total_decode_time} and the Normal distribution assumption from CLT does not apply for a single request, we approximate it using the maximum possible number of output tokens for the model.
As the request queue grows, the waiting time $W_q$ dominates the completion time $C_q$ and the error introduced by the above heuristic reduces as shown in Figure~\ref{fig:base_estimator_accuracy}.
However, for short queues, we maintain the conservative estimate of decode time as it dominates the completion time.

If GPU memory was not a constraint, the decode steps would not be interrupted, and the total decode time would simply be the product of the number of output tokens ($O_q$) and decode time per output token ($d$).
However, LLM serving systems cannot ensure this ideal behavior due to continuous batching.
As requests are added continuously to the GPU's running batch, some requests inevitably exceed the total GPU memory capacity limit and have to be temporarily preempted.
This leads to inefficiency in the generation process that we capture with the inefficiency factor $\epsilon$, i.e., a constant multiplied by the decode time per token that captures the inefficiency associated with the generation process.

Finally, to estimate the completion time of the entire request group (Equation~\ref{eq:group_completion_time}), we need to take the max of all the completion times of individual requests.
\begin{align}
C = \max_q C_q
\label{eq:group_completion_time}
\end{align}

\noindent
\textbf{Offline Profiling.} 
There are two independent profiling steps required for the RWT estimator:
\begin{enumerate*}[label=(\alph*)]
    \item \emph{Workload Profiling:} samples multiple requests from the workload to generate a distribution for input and output tokens, and
    \item \emph{Hardware Profiling:} requires running the model with a single batch of requests on the specific GPU.
    Fixed variables associated with the model and hardware setup, such as the prefill time ($P$), inefficiency factor ($\epsilon$) and decode time per iteration ($d$) are obtained by directly logging these metrics from the LLM serving instance.
    In our implementation, we add these logging metrics directly into vLLM code.
\end{enumerate*} 
\section{Global Scheduler}
\label{s:plan_generation}

The global scheduler is invoked by the RWT estimator when an SLO violation is likely to occur.
Upon invocation, the global scheduler runs a linear programming model to reorder the virtual queues that decide underlying LSO actions to maximize SLO attainment.
The global scheduler uses a linear program solver because it:
\begin{enumerate*}[label=(\alph*)]
    \item allows handling non-determinism by representing request group completion times as distributions, and
    \item offers a systematic way to model various constraints associated with SLOs, model swapping times, and hardware heterogeneity.
\end{enumerate*}
In this section, we present the linear programming model with its defined variables listed in \cref{table:params_variables}.

\begin{table}[!bt]
\caption{Glossary of symbols used in the linear programming solver.}
\resizebox{\linewidth}{!}{
\centering
\begin{tabular}{cl}
\toprule
\textbf{Symbol} & \textbf{Description} \\
\midrule
$g \in \mathbb{G}$ & The $g$-th virtual queue (VQ) in all virtual queues $\mathbb{G}$ \\
$i \in \mathbb{I}$ & The $i$-th request group (RG) in all request groups $\mathbb{I}$ \\
$j$ & Virtual queue position in $[0, L-1]$ with queue length $L$ \\
$x_{g,i,j}$ & Binary decision variable for assignment of RG $i$ to VQ $g$ \\
$wt_{g,j}$ & Request group waiting time \\
$m_{g,j}$ & The model assignment on the $j$-th position of VQ $g$ \\
$t_{g,j}$ & Binary variable for switching the model to serve on VQ $g$ \\
$S$ & Swap time associated with loading a new model into GPU memory \\
$slo_{g,j}$ & SLO preservation rate serving the $j$-th model on VQ $g$ \\
$p_{g,j}$ & Penalty for SLO violation serving the $j$-th model on VQ $g$ \\
\bottomrule
\end{tabular}}
\label{table:params_variables}
\end{table}

\noindent \textbf{Overall Modeling Approach.} 
The goal of the linear programming solver is to find an assignment of request groups to the virtual queues so that all SLOs are met.
To model SLO attainment, we define a \emph{penalty} term for each request group, which is the difference between waiting time and SLO value. 
If SLOs are met, all penalty terms would be smaller than 0.
Given the SLOs as the inputs to the linear programming model, we obtain the request group waiting time estimation from the RWT estimator to estimate the defined penalty terms.
% To estimate penalty terms, we need to model SLOs and request group completion times.
% As SLOs are input to the linear programming model, we do not need to model it, and the other remaining variable is the request group completion time.
% We use the RWT estimator from~\cref{s:base_estimator} to estimate this completion time.
The worst-case waiting time for a request group is the sum of the waiting time for request groups ahead in the virtual queue from the same model (from Equation~\ref{eq:base_equation}), the completion time for the request group for different models (from Equation~\ref{eq:group_completion_time}), and swap times associated with transferring model weights into GPU memory~\footnote{We measure swap times from model load time profiling.}.
Note that effects associated with hardware and model heterogeneity (such as token throughput and eviction vs. swap) that impact request group completion time are captured by the RWT estimator profiling.

\noindent
\textbf{Definitions of Constraints.}
Now, we describe each of the constraints in further detail.
We assume that each virtual queue can have a maximum length, and every request group is assigned to one of the positions in the virtual queue.
Equation~\ref{eq:slot_assignment} models request group assignment to a position in the virtual queue.
\begin{equation}
\sum_{g} \sum_{j} x_{g,i,j} = 1 \forall i \quad
\sum_{i} x_{g,i,j} = 1 \forall g,j
\label{eq:slot_assignment}
\end{equation}

\noindent
Each request group has a one-to-one mapping with a position in a virtual queue.
% is assigned to a single position and vice versa, i.e., each position is assigned to a single request group.
If there are empty positions, we assign them ``empty'' request groups to match request groups and virtual queue capacity.

Each position in the virtual queue would have a corresponding model and SLO based on Equation~\ref{eq:slot_assignment}.
This assignment is captured with Equation~\ref{eq:model_assignment} and Equation~\ref{eq:slo_assignment}.
\begin{equation}
m_{g,j} = \sum_{i} \text{models}_i \times x_{g,i,j} \forall g,j
\label{eq:model_assignment}
\end{equation}
\begin{equation}
    slo_{g,j} = \sum_{i} \text{slos}_i \times x_{g,i,j} \forall g,j
\label{eq:slo_assignment}
\end{equation}

The transition between two different models is captured in Equation~\ref{eq:transition_time}.
While inequalities cannot be directly modeled as constraints, we apply the standard big-M method to reduce the inequality into linear constraints~\cite{cococcioni2021big}.
\begin{equation}
t_{g,j} = (m_{g,j-1} \neq m_{g,j}) \forall g,j
\label{eq:transition_time}
\end{equation}

% Estimating Cumulative Completion Time per GPU Slot & $
The cumulative waiting times of all positions in the virtual queue would be the sum of waiting time, completion times, and swap times as represented in Equation~\ref{eq:completion_time}. $C$ is the completion time of a request group from Equation~\ref{eq:group_completion_time}.

\begin{equation}
\text{wt}_{g,j} = \sum_{i} \sum_{k}^{j-1} W_{g,i} \times x_{g,i,k} + \sum_{k}^{j-1} t_{g,k} \times S + \sum_{i} \sum_{k}^{j-1} C_{g,i} \times t_{g,k} \times x_{g,i,j} \forall g,j 
\label{eq:completion_time}
\end{equation}

\noindent

% Estimating Penalty for Violating SLOs
The penalty would simply be the difference between the waiting time and the SLO value, as shown in Equation~\ref{eq:penalty}.
\begin{equation}
p_{g,j} = wt_{g,j} - slo_{g,j} \forall g,j
\label{eq:penalty}
\end{equation}

% Estimating Penalty for Violating SLOs
The final constraint is that all penalty values should be less than 0 i.e., all SLOs are satisfied.
\begin{equation}
    p_{g,j} \leq 0 \forall g,j
\end{equation}

\noindent
\textbf{Optimization Goal.}
The linear programming model aims to minimize the total penalty for SLO violations.

\begin{equation}
\min (\sum_{g} \sum_{j} p_{g,j})
\end{equation}
\section{Evaluation}
\label{s:results}

%Add autoscaling with QLM line and describe additional resources needed.

Our experiments address the following research questions:
\begin{enumerate}[label=(\alph*)]
    \item \sysname performance with respect to SLO attainment and request throughput in single-model serving (\cref{ss:single_model_eval}),
    \item \sysname performance with respect to SLO attainment and request throughput in multi-model serving (\cref{ss:multi-model-eval}),
    \item Contribution of each LSO to \sysname performance,
    \item Accuracy of the RWT estimator in request waiting time estimation,
    \item Robustness analysis of \sysname to hardware heterogeneity, token distributions, burstiness, and request group size regarding LLM-serving performance (\cref{ss:robustness}), and
    \item Overhead of using \sysname with increasing queue sizes.
\end{enumerate}

\noindent \textbf{Experiment Setup.}
We evaluate \sysname on multiple varying-sized open-source LLMs: Mistral-7B~\cite{mistral}, Vicuna-13B~\cite{vicuna}, and Llama-70B~\cite{touvron2023llama}.
We evaluate on a test bed consisting of GPUs of two types: 30 NVIDIA A10 (24 GB memory) and 50 NVIDIA A100 (80 GB memory).
The setup represents both model and device heterogeneity.
To evaluate the benefit of \sysname, we consider the following three baseline mechanisms:
\begin{enumerate*}[label=(\arabic*)]
    \item \textit{EDF} (Earliest Deadline First): Requests are sorted by their SLO values such that requests with the smallest SLO values are at the front of the virtual queue,
    \item \textit{vLLM}~\cite{kwon2023efficient}: Requests use the default first-come-first-serve (FCFS) scheduler in vLLM, and
    \item \textit{SHEPHERD}~\cite{zhang2023shepherd}: Requests are served with dynamic batching and an ILP formulation for ordering and placement.
    Note that SHEPHERD cannot be easily extended to work with continuous batching as the LP formulation assumes fixed batches with deterministic execution times.
\end{enumerate*}

%although we normalize we explain all results wrt to vLLM industry standard
%explain how different models are impossible to explore, so we choose production requirements
%replace interactive with latency sensitive
\begin{figure}[!h]
\centering
    \includegraphics{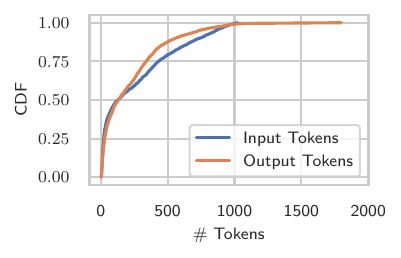}
    \caption{Distribution of input and output tokens in the shareGPT dataset.}
    \Description{Distribution of input and output tokens in the shareGPT dataset.}
    \label{fig:token_distribution}
\end{figure}

\noindent
\textbf{Workloads.}
We create our experimental workloads from the requirements of a production cloud service provider except for request arrival rates due to confidentiality reasons.
Request arrivals are modeled with a Poisson distribution and queues are created by varying the arrival rates.
Each workload trace uses 3,500 requests from the ShareGPT~\cite{shareGPT} dataset with input/output token distribution as shown in Figure~\ref{fig:token_distribution}.
We classify all requests into three categories and define their SLO values accordingly: 
\begin{enumerate*}
    \item \emph{Interactive:} 20s,
    \item \emph{Batch-1:} 1 min, and
    \item \emph{Batch-2:} 1 hour.     
\end{enumerate*}
Note that these SLOs are defined with respect to the 99th percentile value of the time to first token (TTFT).
% While our choice of SLO values is motivated by production requirements, we also provide robustness analysis for alternative SLO values in Appendix~\ref{s:slo_value_sweep}.
We test the experimental workload in the following three scenarios:
\begin{enumerate*}[label=\textbf{[$\mathbf{W_\Alph*}$]}]
    \item \textbf{\textit{Single-Model Interactive and Batch Workload}} which consists of Batch-1, Batch-2, and Interactive requests for a single model, and no model swapping is required.
    \item \textbf{\textit{Multi-Model Batch Workload}} which consists of Batch-1 and Batch-2 requests.
    Batch-1 requests use two models: fine-tuned versions of Mistral 7B and Llama 70B.
    Batch-2 requests use three models: fine tuned versions of Vicuna 13B and Llama 70B.
    \item \textbf{\textit{Single-Model MegaPrompt Workload}} This request workload consists of several ``mega prompts'' in addition to the workload from $W_B$.
    To generate the mega prompt workload, we randomly select requests with total input and output tokens in the 3K -- 4K range.
    These mega prompts have a large number of input and output tokens that occupy a large percentage of GPU memory and cause further HOL blocking.
\end{enumerate*}

We do not present results for the multi-model interactive workload because \sysname assigns a separate GPU for each model, effectively reducing the workload to a single model workload ($W_A$). 
\sysname' global scheduler decides not to swap the models as the model swapping time exceeds the interactive request SLO (20s).
\newline
\begin{figure*}[!t]
    \centering    
    \begin{minipage}[t]{.32\textwidth}
        \centering
        \includegraphics{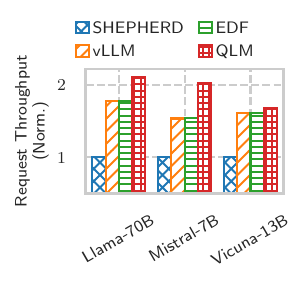}
        \caption{Single model request serving throughput at 0.5K requests/s interactive arrival rate. Increased throughput corresponds to 1.1-2.3\texttimes{} GPU requirement reduction.}
        \Description{Single model request serving throughput at 0.5K requests/s interactive arrival rate. Increased throughput corresponds to 1.1-2.3\texttimes{} GPU requirement reduction.}
        \label{fig:single_model_throughput}
    \end{minipage}%
    \hfill
    \centering 
    \begin{minipage}[t]{.32\textwidth}
        \centering
        \includegraphics{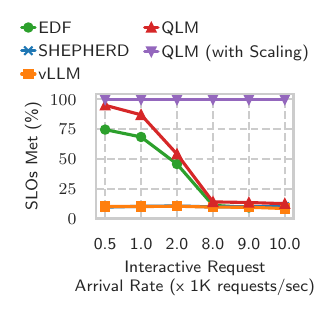}
        \caption{Single model SLO satisfaction for varying interactive request arrival rates for Vicuna 13B.}
        \Description{Single model SLO satisfaction for varying interactive request arrival rates for Vicuna 13B.}
        \label{fig:single_model_SLOs}
    \end{minipage}%
    \hfill
    \centering    
    \begin{minipage}[t]{.32\textwidth}
        \centering
        \includegraphics{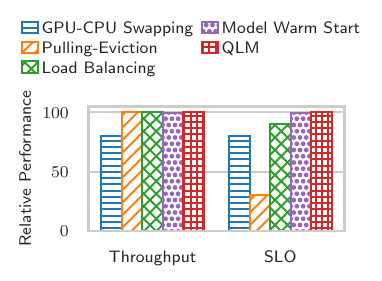}
        \caption{Single model LSO ablation study at 0.5K requests/s interactive arrival rate for Vicuna 13B.}
        \Description{Single model LSO ablation study at 0.5K requests/s interactive arrival rate for Vicuna 13B.}
        \label{fig:single_model_ablation}
    \end{minipage}
    \hfill
\end{figure*}

\subsection{Single-Model Evaluation}
\label{ss:single_model_eval}

\BLUE{We run workload $W_A$ on 50 A100 GPUs to evaluate the \textit{single-model} LLM serving performance regarding the request throughput, SLO attainment, and LSO contribution ablation study (similar to the multi-model evaluation in \cref{ss:multi-model-eval}).}

\noindent
\textbf{Request Throughput and SLO Attainment.}
Figure~\ref{fig:single_model_SLOs} shows the percentage of SLOs that are satisfied by \sysname and the baseline systems.
We find that when the request arrival rate significantly exceeds the serving capacity, none of the systems can satisfy the SLOs.
This is because the minimum serving time is much longer than the specified SLO.
As the arrival rate of interactive requests decreases, \sysname performs the best in satisfying the maximum number of SLOs.
Specifically, it performs better than the baseline mechanisms because:
\begin{enumerate*}[label=(\alph*)]
    \item Compared to vLLM, \sysname is able to move interactive service requests ahead in the queue,
    \item Compared to EDF, \sysname enables appropriate eviction of batch requests from the running queue, and
    \item Compared to SHEPHERD, \sysname uses continuous batching as opposed to static batch size and models the auto-regressive LLM nature with the RWT estimator to increase request throughput.
\end{enumerate*}

We find that the advantages of \sysname with respect to smart selection among various LSOs, continuous batching, and appropriate request prioritization help with improving request throughput.
Figure~\ref{fig:single_model_throughput} shows the request throughput for \sysname and the individual baseline mechanisms at arrival rate of 0.5K requests/s where \sysname is able to achieve the maximum SLO satisfaction.
\sysname achieves higher throughput, i.e., 20\% higher compared to vLLM and EDF, and 50\% higher than SHEPHERD.

\noindent
\textbf{Contribution of Each LSO.}
Figure~\ref{fig:single_model_ablation} shows the impact of removing each LSO considered by the backend LLM serving instance in \sysname.
Request pulling contribute significantly to latency reduction for interactive services and consequently increase the number of SLOs met.
Request eviction increases request throughput by swapping the KV cache into CPU memory.
Finally, model swapping has no impact on this workload as a single model is being served.

\begin{figure*}[!t]
    \centering    
    \begin{minipage}[t]{.32\textwidth}
        \centering
        \includegraphics{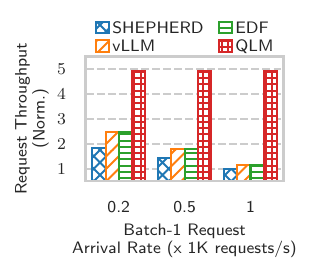}
        \caption{Multi-model request serving throughput for varying Batch-1 request arrival rates. Increased throughput corresponds to 2-5\texttimes{} GPU requirement reduction.}
        \Description{Multi-model request serving throughput for varying Batch-1 request arrival rates. Increased throughput corresponds to 2-5\texttimes{} GPU requirement reduction.}
        \label{fig:multi_model_throughput}
    \end{minipage}%
    \hfill
    \centering
    \begin{minipage}[t]{.32\textwidth}
        \centering
        \includegraphics{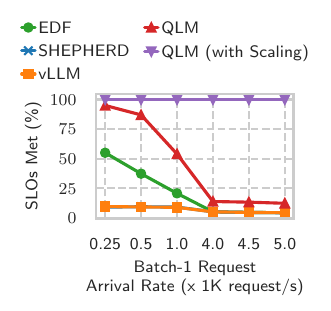}
        \caption{Multi-model SLO satisfaction for varying Batch-1 request arrival rates.}
        \Description{Multi-model SLO satisfaction for varying Batch-1 request arrival rates.}
        \label{fig:multi_model_slos}
    \end{minipage}%
    \hfill
    \centering 
    \begin{minipage}[t]{.32\textwidth}
        \centering
        \includegraphics{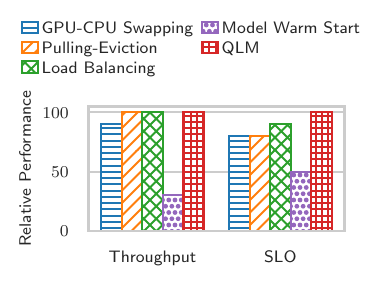}
        \caption{Multi-model LSO ablation study for 0.25K requests/sec Batch-1 arrival rate.}
        \Description{Multi-model LSO ablation study for 0.25K requests/sec Batch-1 arrival rate.}
        \label{fig:multi_model_ablation}
    \end{minipage}%
    \hfill
\end{figure*}

\subsection{Multi-Model Evaluation}
\label{ss:multi-model-eval}
\BLUE{We run workload $W_B$ to evaluate the \textit{multi-model} performance on 50 A100 GPUs with respect to request throughput and SLO satisfaction.}
Additionally, we also provide an ablation study to understand the contribution of each LSO to the overall \sysname performance.

\noindent
\textbf{Request Throughput and SLO Attainment.}
Figure~\ref{fig:multi_model_throughput} shows the request throughput (i.e., requests served per second) for $W_B$ comparing \sysname with the baseline mechanisms for varying Batch-1 arrival rates.
\sysname provides up to 3--4\texttimes{} higher throughput due to the following factors:
\begin{enumerate*}[label=(\arabic*)]
    \item The use of request groups minimizes repeated swapping required as the model would only be swapped in once per request group instead of per individual request, and
    \item The global scheduler couples every Batch-1 model with another Batch-2 model to minimize swaps while maintaining an equal distribution of queue sizes.
\end{enumerate*}

The improvement in request throughput directly maximizes the percentage of SLO satisfied for all requests.
Figure~\ref{fig:multi_model_slos} shows the percentage of SLO satisfied for the interactive services against against their arrival rate.
When the Batch-1 requests arrive at less than 0.5K requests/s, \sysname satisfies more than 90\% of all SLO values.
As the arrival rate of Batch-1 service requests increases, no combination of requests would be able to meet all SLOs, and the global scheduler would fail to return a solution.
In such a scenario, a scale-up action is required to add more GPU devices.
We perform this scale-up action to enable 100\% SLO attainment if the current GPU capacity is insufficient. 
The baselines perform worse compared to \sysname because none of them consider the impact of model swapping.
Other limitations of the baselines are discussed in Section~\ref{ss:single_model_eval}.

\noindent
\textbf{Contribution of Each LSO.}
Each of the four LSOs used by \sysname, including request pulling, request eviction, model swapping, and load balancing, contributes to either the latency and/or the throughput of the serving system.
Figure~\ref{fig:multi_model_ablation} shows the impact of removing each LSO on \sysname performance for $W_B$.
The model warm start LSO contributes the most to \sysname performance for both SLOs and throughput, as multiple models need to be multiplexed on the same LLM serving instance.
Additionally, the other LSOs contribute primarily to the latency SLO attainment.

\subsection{\sysname Robustness Analysis}
\label{ss:robustness}

\begin{figure*}[!t]
    \centering
    \hspace{-4ex}
    \begin{minipage}[t]{.32\textwidth}
        \centering
                                    \includegraphics{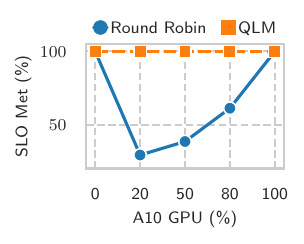}
        \caption{Impact of hardware heterogeneity.}
        \Description{Impact of hardware heterogeneity.}
        \label{fig:hardware_heterogeneity}
    \end{minipage}%
    \centering 
    \hspace{0.5ex}
    %fig here
        \begin{minipage}[t]{.32\textwidth}
        \centering
        \includegraphics{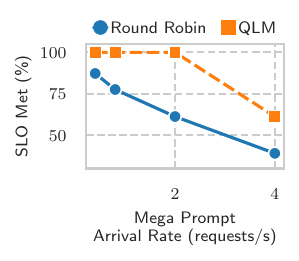}
        \caption{Impact of mega prompt arrivals.}
        \Description{Impact of mega prompt arrivals.}
        \label{fig:mega_distribution}
    \end{minipage}%
    \centering
    \hspace{0.5ex}
    \begin{minipage}[t]{.32\textwidth}
        \centering
        \includegraphics{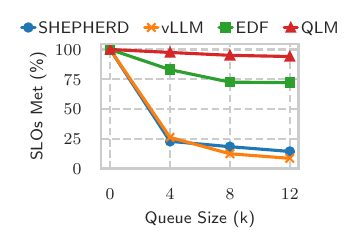}
        \caption{Impact of increasing queue size on SLO satisfaction.}
        \Description{Impact of increasing queue size on SLO satisfaction.}
        \label{fig:varying_burstiness}
    \end{minipage}%
\end{figure*}

\noindent
\textbf{Hardware Heterogeneity.}
% \sysname is able to outperform other baseline mechanisms when the GPU hardware is heterogeneous, i.e.,
% the nature of GPUs is different across the cluster.
%ablation study of load balancer on herterogenous hardware
We run $W_A$ on a mix of A10 and A100 GPUs to evaluate the robustness of \sysname performance in heterogeneous hardware setup.
Figure~\ref{fig:hardware_heterogeneity} shows request throughput when the cluster has varying ratios of A10 to A100 GPUs.
The A10 is a lower-end GPU with $\sim$3\texttimes{} lower GPU memory and thus is only capable of serving a much lower request throughput compared to the A100 GPU.
\sysname takes into account this difference between request throughput across GPUs via the RWT estimator with offline profiling, and the global scheduler proportionally assigns a lower number of requests to the A10 GPU compared to the A100.
On the other hand, if we use a round-robin policy for request assignment to the LLM serving instances (while using default \sysname policy per instance), the load would be distributed equally, leading to higher queue drain times for the A10 GPU.
Additionally, we also observe that the benefit of \sysname is more compared to a random policy when the heterogeneity of the cluster is higher.
When the A10 GPUs constitute 20--50\% of the cluster (more heterogeneous), the improvement of \sysname over random policy is 2--5\texttimes{} higher compared to a 100\% A10 or 100\% A100 composition (more homogeneous).
Note that for experimental purposes, we increase the total number of A100s proportionately to demonstrate the impact of GPU imbalance. 

\noindent
\textbf{Mega Prompt Workload.}
The RWT estimator of \sysname takes into account input and output token distribution when estimating the request waiting time.
Consequently, when there are distinct token distributions, such as in workload setup $W_C$, \sysname is able to load balance them intelligently across LLM serving instances to minimize the queue drain time.
For example, in workload $W_C$, the ``mega prompts'' use a large number of tokens, and their KV cache occupies the entire GPU memory, causing head-of-the-line blocking for the regular requests in the queue.
The optimal policy, as identified by \sysname, in such a scenario would be to allocate all the regular requests to another LLM serving instance.
Note that request eviction is not an option if all SLOs are tight.
Figure~\ref{fig:mega_distribution} shows the benefit of \sysname for workload $W_C$.
The relative benefit of \sysname is highest for a few mega prompts because the regular requests can be moved to another GPU.
As the percentage of mega prompts increases, there is no option but to assign them to different LLM serving instances, causing inevitable HOL blocking, and the benefit of \sysname reduces.
In such a case, we would need to perform a scale-up action and add more GPU devices to the cluster to continue maintaining SLOs.

\noindent
\textbf{Varying Queue Size and Burstiness.}
The benefit of \sysname is largely present when the queue size is large, and intelligent decision-making is required for setting LSO actions.
Thus, to show the benefit of \sysname under varying queue sizes, we vary the arrival rates of requests in $W_B$ to create a large queue and compare it against the baseline systems as shown in Figure~\ref{fig:varying_burstiness}.
When the queue size is 0, \sysname offers no benefit in maintaining SLOs as compared to the baseline approaches because the system is underutilized and does not require any smart decision-making.
However, as the queue size increases, the percentage of SLOs met by the baseline systems keeps dropping due to reasons described in Section~\ref{ss:single_model_eval}, while \sysname is able to maintain a high SLO satisfaction percentage.

\begin{figure*}[!t]
    \centering
        \begin{minipage}[t]{.32\textwidth}
        \centering
        \includegraphics{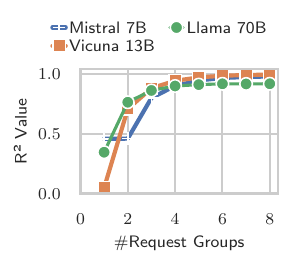}
        \caption{Accuracy of RWT estimator.}
        \Description{Accuracy of RWT estimator.}
        \label{fig:base_estimator_accuracy}
    \end{minipage}%
    \hfill
    \centering
    \begin{minipage}[t]{.34\textwidth}
    \centering    \includegraphics{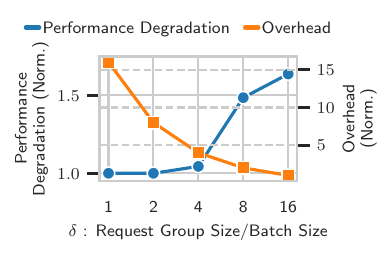}
    \caption{Impact of request group size on \sysname performance.}
    \Description{Impact of request group size on \sysname performance.}
    \label{fig:delta_variation}
    \end{minipage}%
    \hfill
    \begin{minipage}[t]{.3\textwidth}
        \centering
        \includegraphics{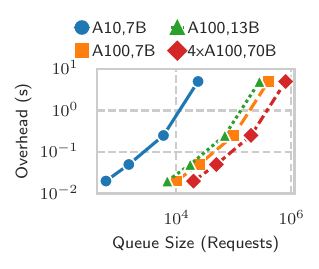}
        \caption{\sysname Overhead.}
        \Description{\sysname Overhead.}
        \label{fig:overhead}
    \end{minipage}%
\end{figure*}

%GPU memory utilization (QLM does less work in single model case)

\noindent
\textbf{RWT Estimator Accuracy.}
\BLUE{The RWT estimator calculates the request waiting time based on initial profiling of the model and hardware setup.
This initial profiling time is negligible as only a single batch of requests needs to be run on the GPU.

% As described in Section~\ref{s:base_estimator}, \sysname does not assume that the exact output tokens are known ahead of time, but instead uses the workload output token distribution.
Figure~\ref{fig:base_estimator_accuracy} shows the coefficient of determination ($R^2$ values) when estimating waiting times for increasing queue sizes across the different models.
Overall, we confirm our observation that as the number of request groups in the queue increase the waiting time estimation becomes more accurate.
Specifically, we find with four request groups the RWT estimator reaches an accuracy of 0.99.

While the RWT estimator is highly accurate in estimating request waiting time for longer request queues, it is not perfect.
When request queues are small, statistical averaging effects of continuous batching do not hold and the waiting time estimation tends calculates more conservative (i.e. higher) estimates leading to lower estimation accuracy.
}

\noindent
\textbf{Impact of Request Group Size.}
\sysname sets the request group size as a multiple ($\delta$) of the average batch size.
The exact $\delta$ value depends on the acceptable trade-off between the overhead of running the global scheduler and the granularity of decision-making.
As $\delta$ becomes smaller, \sysname achieves a finer granularity of decision-making, leading to improved performance.
However, the overhead leads to delayed decision-making.
Figure~\ref{fig:delta_variation} demonstrates this tradeoff between performance degradation (caused by changing granularity in decision making) and overhead of the global scheduler when varying $\delta$.
At $\delta=16$, the overhead is smallest, but decision-making granularity is coarse, leading to sub-optimal decisions (such as imbalance between virtual queue sizes of LLM serving instances).
In contrast, at $\delta=1$, the performance degradation is minimal, but overhead is much higher.
We choose $\delta=4$, as it results in nearly zero performance degradation, compared to $\delta=1$, while maintaining a low overhead.

\noindent
\textbf{Scalability and Overhead.}
The global scheduler is invoked only when \sysname detects an SLO violation using the RWT estimator.
While the global scheduler is performing the virtual queue reordering, \sysname continues serving and requests with tight SLOs (low latency requirements) placed at the front of the virtual queues are not interrupted.
This ensures the global scheduler is off the critical serving path (i.e. the overheads can be hidden) and ensures SLO compliance.
While the global scheduler is the primary overhead, other system overheads include the CPU memory required for the request eviction and model swapping LSOs.
Specifically, we require an additional 80 GB CPU memory for Vicuna 13B and Mistral 7B, and 320 GB CPU memory for Llama 70B.

In Figure~\ref{fig:overhead}, we show the time required to solve for the linear program in the global scheduler with varying queue sizes in terms of the number of requests.
As the basic unit of the solver is a single request group, the model and GPU configurations with a larger request group size would be able to handle a much larger queue size for the same overhead.
Consequently, configurations with a large request group size, such as an A100 with a 7B model, can handle a maximum queue size of 400K requests at a 5s overhead per request group (i.e., 5 ms per request, a production requirement).

%1. R^2 value for varying queue sizes (smaller queue should show worse R^2)
%2. SLO, arrival values from the jupyter,
%3. Lower SLO for Fig 9, Fig 12 
\section{Discussion and Future Work}
\label{s:discussion}

\noindent \textbf{What are the failure scenarios for the RWT Estimator?}
The RWT Estimator has the following drawbacks:
\begin{enumerate*}[label=(\alph*)]
\item If the number of request groups in the queue are small, \sysname overestimates the request waiting time. Such a conservative estimate is important as it would still lead to SLO preservation by the global scheduler, albeit the system could be underutilized compared to optimal.
\item When the number of output tokens is out-of-distribution (for example, more than 5-10 standard deviations above mean), the estimator would underestimate the request group completion time leading to potential SLO violations. However, in our experience with shareGPT and production traces such scenarios are rare.
\end{enumerate*}

\noindent \textbf{What happens if \sysname is unable to meet SLOs?}
%business question
\sysname's global scheduler may not be able to find an optimal virtual queue ordering if the request demand is high and the number of LLM serving instances (i.e., underlying compute resources) is insufficient.
In such cases, we have three choices:
\begin{enumerate*}[label=(\alph*)]
    \item scale up the number of LLM serving instances by adding GPU devices, as we demonstrate in Figure~\ref{fig:multi_model_slos} and Figure~\ref{fig:single_model_SLOs}, 
    \item fall back to a heuristic such as Earliest Deadline First (EDF) and continue serving requests, and
    \item performance admission control or rate limiting by dropping incoming requests to limit queue size.
\end{enumerate*}
Option (a) can only be used when there are available resources, whereas Option (b) and Option (c) would lead to SLO violations.

\noindent \textbf{Can new LSOs be added to \sysname?}
\sysname can be extended to support other LSOs that depend on the queue size and request ordering.
For example, GPU partitioning techniques (such as NVIDIA MIG~\cite{mig}) can be added as an LSO with additional constraints on memory in the linear programming solver described in \cref{s:plan_generation}.
We leave the addition of extra LSOs to our future work.

\noindent \textbf{How can \sysname handle request priorities?}
\sysname can also be used when strict priorities are assigned to requests instead of SLO values.
In the strict priority model, $request_1$ would execute before $request_2$, if $priority(request_1)$ is less than  $priority(request_2)$.
With strict priority, the relative ordering of requests across priorities is pre-decided, however per-priority assignment still needs to be optimized to maximize SLO satisfaction.
Consequently, the concepts of virtual queues, request groups, and the RWT estimator continue to remain useful. 

\noindent \textbf{Can SLOs be defined on end-to-end latency?}
\BLUE{\sysname addresses the problem of optimizing time to first token (TTFT) SLO attainment.
However, it can also be made to work with the modified objective of end to end request completion time SLOs (i.e. including decoding iterations).
In this case, the RWT estimator and global scheduler need to be modified to consider a distribution of number of output tokens per request (as the exact number are unknown aprioiri).}

\noindent \textbf{Can SLOs be defined on inter-token latency?}
\sysname primarily addresses the problem of TTFT and does not guarantee the inter-token latency (ITL).
However, related work such as Andes~\cite{liu2024andes} attempts to guarantee token generation speed (that can be converted into inter-token latency) by modifying the local vLLM scheduler.
\sysname can work together with such systems to ensure ITL SLOs along with TTFT SLOs.

\section{Related Work}
\label{s:related_work}

\setlength{\tabcolsep}{10pt}

\begin{table*}[!ht]
\centering
\small
\renewcommand{\arraystretch}{1.8}  % Add this line to increase row spacing
\begin{threeparttable}
\begin{tabular}{lcccc}
\Xhline{1.2pt}
\textbf{Category} & \textbf{Multi Instance} & \textbf{SLO-aware} & \parbox[c]{2.2cm}{\centering \textbf{Autoregressive} \\ \textbf{Optimizations}} & \parbox[c]{2.2cm}{\centering \textbf{Hardware} \\ \textbf{Heterogeneity}} \\ \Xhline{1.2pt}
\parbox[c]{5cm}{LLM Scheduling Optimizations~\cite{yu2022orca,kwon2023efficient,tgi}} & \xmark & \xmark & \cmark & \cmark \\ \hline
\parbox[c]{5cm}{LLM Serving Backend Optimizations\\~\cite{liu2023deja,sheng2023flexgen,llm-inference,zhang2023h2o,liu2023scissorhands,ge2024model,miao2023spotserve,abhyankar2024apiserve,zhong2024distserve,patel2023splitwise,sheng2023slora,zhu2024relayattention,fang2021turbotransformers,leviathan2023fast}} & \xmark & \xmark & \cmark & \cmark \\ \hline
\parbox[c]{5cm}{General ML Model-Serving Systems\\~\cite{crankshaw2017clipper,zhang2019mark,inferline,zhang2023shepherd,gujarati2020serving}} & \cmark & \cmark & \xmark & \cmark \\ \hline
\parbox[c]{5cm}{LLM Orchestration Systems~\cite{sun2024llumnix}} & \cmark & \xmark & \cmark & \xmark \\ \Xhline{1.2pt}
\textbf{\sysname} & \cmark & \cmark & \cmark & \cmark \\ \Xhline{1.2pt}
\end{tabular}
\begin{tablenotes}
\footnotesize
\item \cmark: Supported, \xmark: Not Supported
\end{tablenotes}
\caption{Comparison of \sysname with Related Work}
\end{threeparttable}
\end{table*}

\noindent \textbf{LLM Scheduling and Orchestration.}
Existing state-of-the-art LLM serving systems~\cite{yu2022orca,kwon2023efficient,tgi} adopts continuous batching and a first-come-first-serve (FCFS) scheduling policy that suffers from head-of-line (HOL) blocking, which we address in \sysname.
FastServe~\cite{wu2023fast} proposes preemptive scheduling with a Multi-Level Feedback Queue. 
Andes~\cite{liu2024andes} defines Quality-of-Experience (QoE) for LLM serving as token delivery speed, and proposes a preemptive scheduler that maximizes QoE.
Llumnix~\cite{sun2024llumnix} is an orchestration system across multiple LLM serving instances.
\sysname is the first queue management framework that optimizes SLO attainment while improving LLM-serving throughput and device utilization by systematically orchestrating backend LSOs.

\balance
\noindent \textbf{LLM Serving Backend Optimization.}
Various LLM serving backend optimization techniques have been proposed to improve token generation throughput and memory cost while adapting to fine-tuning paradigms such StreamingLLM, Speculative Decoding, ChunkedAttention, FlashAttention and more~\cite{liu2023deja,sheng2023flexgen,llm-inference,zhang2023h2o,liu2023scissorhands,ge2024model,miao2023spotserve,abhyankar2024apiserve,zhong2024distserve,patel2023splitwise,sheng2023slora,zhu2024relayattention,fang2021turbotransformers,leviathan2023fast}.
These backend LLM-serving optimizations are complementary to \sysname as the LLM serving instance (see \cref{def:serving-instance}), and their impact on token generation throughput can be captured with profiling for the RCT Estimator (see \cref{s:base_estimator}).

\noindent \textbf{General ML Model-Serving Systems.}
Traditional model-serving systems provide functionalities such as scheduling, placement, batching, and autoscaling.
Clipper~\cite{crankshaw2017clipper}, TensorFlow-Serving~\cite{tensorflow-serving}, MArk~\cite{zhang2019mark}, InferLine~\cite{inferline}, SHEPHERD~\cite{zhang2023shepherd}, and Clockwork~\cite{gujarati2020serving} are some earlier work on serving traditional ML models like ResNet that are relatively small.
INFaaS~\cite{romero2021infaas} and Cocktail~\cite{gunasekaran2022cocktail} propose a model-less serving framework to automate the model selection and autoscaling to meet SLOs.
However, they fail to consider the autoregressive property of LLMs.
On the other hand, advanced autoscaling techniques are complementary to \sysname. 
\section{Conclusion}
\label{s:conclusion}

We presented \sysname, a novel queue management framework that orchestrates backend LSOs for SLO-oriented LLM serving.
Evaluation using real-world LLM serving datasets on heterogeneous model types and GPU devices demonstrate that \sysname improves end-to-end latency SLO attainment by 40--90\% while improving serving throughput and device utilization by 20-400\%.
% We further discuss extensions to \sysname in Appendix~\ref{s:discussion}.

% We demonstrated the use of an intelligent decision engine to orchestrate backend LSOs through \sysname. 
% In the future, we plan to enhance \sysname in several ways:
% \begin{enumerate*}[label=(\roman*)]
%     \item \textit{incorporate additional LSOs} such as auto-scaling, MIG partitioning, and configuration optimization of inferencing servers.
%     \item \textit{improve fault tolerance and scalability} by adopting principles of distributed system design.
%     \item \textit{accommodate emerging models and deployment options} such as s-LoRA and RAG.
% \end{enumerate*}
\section{Acknowledgements}
\label{s:acknowledgements}

We thank the anonymous reviewers for providing their valuable feedback.
We also thank Nick Hill, Lionel Villard, and Mudhakar Srivatsa for providing us with production requirements and insight, and Rohan Arora for infrastructure support.
This work is supported by the National Science Foundation (NSF) under grant No. CCF 20-29049 and by the IBM-ILLINOIS Discovery Accelerator Institute (IIDAI).
Any opinions, findings, conclusions, or recommendations expressed in this material are those of the authors and do not necessarily reflect the views of the NSF or IBM.

\bibliographystyle{ACM-Reference-Format}
\bibliography{socc}
\appendix
\section{Appendix}

\subsection{Waiting Time Predictability}
\label{s:waiting_time_proof}

\begin{table}[h]
\centering

\label{tab:symbols}
\begin{tabular}{c|c}
\hline
\textbf{Symbol} & \textbf{Description} \\
\hline
$W$ & Total waiting time for a request in the queue \\
$O$ & Total number of output tokens in the queue \\
$\Theta$ &  Output token generation throughput (tokens/s) \\
$B$ & Batch size  \\
$\delta$ & Decoding time per token \\
$\epsilon$ & Preemption factor \\
$\text{GPU}$ & Total token memory capacity in GPU\\
$I_i$ & Number of input tokens for the $i$-th request \\
$O_i$ & Number of output tokens for the $i$-th request \\
$\mu_O$ & Mean number of output tokens per request \\
$\sigma_O^2$ & Variance of output tokens per request \\
$\mu_I$ & Mean number of input tokens per request \\
$\sigma_I^2$ & Variance of input tokens per request \\
n & Size of the waiting queue \\
\hline
\end{tabular}
\caption{Glossary of Symbols Used in the Statistical Derivation}
\end{table}

Below, we present a statistical derivation for the underlying assumptions from the RWT estimator from Section~\ref{s:base_estimator}.

The total number of output tokens in the queue would be the sum of output tokens of individual requests.
The total waiting time for a request joining such a queue would be the time to process output tokens for requests ahead in the queue, which would be the number of output tokens divided by the token generation throughput.

\begin{equation}
W = \frac{O}{\Theta}
\end{equation}

The average token generation throughput (i.e. the number of output tokens generated per second) is simply the average batch size  divided by time to generate each token.

\begin{equation}
    \Theta = \frac{B}{\delta \times \epsilon}
\end{equation}

Due to continuous batching, the batch size is simply the number of tokens that can be kept in GPU memory.

\begin{equation}
  B \approx \frac{\text{GPU}}{I_i + O_i}  
\end{equation}

Simplifying the above equations, we derive the waiting time to be:

\begin{equation}
    W = \frac{O \times \delta \times \epsilon \times E[I_i + O_i]}{GPU}
\end{equation}

Therefore the expected waiting time would be:
\begin{equation}
    E[W] = \frac{E[O] \times \delta \times \epsilon \times E[I_i + O_i]}{GPU}
\end{equation}

As each request is independent of others and all requests originate from the same model, we assume the number of tokens for requests in the batch and queue to be i.i.d. (independent and identically distributed) variables.

Due to the i.i.d. condition, for a large queue, we can apply the Central Limit Theorem and approximate the total number of output tokens with a normal distribution.

\begin{equation}
    O = \mathcal{N}(n\mu_{O}, n \sigma_{O}^2)
\end{equation}

\begin{equation}
    E[O] = n \mu_{O}
\end{equation}

Similarly for a large batch, we can also apply the Central Limit Theorem to approximate the number tokens in a batch with a normal distribution.

\begin{equation}
    I_i + O_i = \mathcal{N}(\mu_{I + O}, \sigma_{I+O}^2/B)
\end{equation}

% Relating Token Generation Throughput and Batch Size
\begin{equation}
    E[I_i + O_i] = \mu_I + \mu_O
\end{equation}

As remaining terms in the waiting time estimation are constant, we find that the expected waiting time in the queue can be estimated analytically.

\end{document}